\documentclass[aps,prd,twocolumn,superscriptaddress, nofootinbib]{revtex4-2}

\usepackage{graphicx}
\usepackage{xcolor}
\usepackage[colorlinks=true,linkcolor=blue,citecolor=blue,urlcolor=blue]{hyperref}
\usepackage{amsmath}
\usepackage{orcidlink}

\begin{document}


\title{High‑Stability Deformable Mirrors for Correcting Non‑Axisymmetric Residual Aberrations in Thermal Compensation of Future Gravitational Wave Interferometers}


\author{L. A. Corubolo\,\orcidlink{0009-0001-5494-3309}}
\affiliation{INFN, Sezione di Roma “Tor Vergata”, I-00133 Roma, Italy}
\affiliation{Università degli Studi di Roma “Tor Vergata”, I-00133 Roma, Italy}

\author{M. Lorenzini\,\orcidlink{0000-0002-2765-7905}}
\affiliation{INFN, Sezione di Roma “Tor Vergata”, I-00133 Roma, Italy}
\affiliation{Università degli Studi di Roma “Tor Vergata”, I-00133 Roma, Italy}

\author{L. Aiello\,\orcidlink{0000-0003-2771-8816}}
\affiliation{INFN, Sezione di Roma “Tor Vergata”, I-00133 Roma, Italy}
\affiliation{Università degli Studi di Roma “Tor Vergata”, I-00133 Roma, Italy}

\author{E. Cesarini\,\orcidlink{0000-0001-9127-3167}}
\affiliation{INFN, Sezione di Roma “Tor Vergata”, I-00133 Roma, Italy}

\author{M. Cifaldi\,\orcidlink{0009-0007-1566-7093}}
\affiliation{INFN, Sezione di Roma “Tor Vergata”, I-00133 Roma, Italy}

\author{M. Ianni\,\orcidlink{0009-0002-0821-2692}}
\affiliation{INFN, Sezione di Roma “Tor Vergata”, I-00133 Roma, Italy}
\affiliation{Università degli Studi di Roma “Tor Vergata”, I-00133 Roma, Italy}

\author{D. Lumaca\,\orcidlink{0000-0002-3628-1591}}
\affiliation{INFN, Sezione di Roma “Tor Vergata”, I-00133 Roma, Italy}

\author{Y. Minenkov\,\orcidlink{0000-0003-2023-6400}}
\affiliation{INFN, Sezione di Roma “Tor Vergata”, I-00133 Roma, Italy}

\author{I. Nardecchia\,\orcidlink{0000-0001-5558-2595}}
\affiliation{INFN, Sezione di Roma “Tor Vergata”, I-00133 Roma, Italy}

\author{A. Rocchi\,\orcidlink{0000-0002-1382-9016}}
\affiliation{INFN, Sezione di Roma “Tor Vergata”, I-00133 Roma, Italy}

\author{C. Taranto\,\orcidlink{000-0003-0431-3875}}
\affiliation{INFN, Sezione di Roma “Tor Vergata”, I-00133 Roma, Italy}

\author{V. Fafone\,\orcidlink{0000-0003-1314-1622}}
\affiliation{INFN, Sezione di Roma “Tor Vergata”, I-00133 Roma, Italy}
\affiliation{Università degli Studi di Roma “Tor Vergata”, I-00133 Roma, Italy}




\begin{abstract}
In gravitational wave detectors, optical aberrations arise mainly from laser absorption in coatings and production process defects in the optics along the laser path. If left uncorrected, these optical path distortions drive the interferometer away from its optimal working point, degrading both stability and sensitivity. Future instruments such as the Einstein Telescope high-frequency detector will operate with unprecedented circulating power, further amplifying the aberration budget.

In the current detectors Advanced Virgo and Advanced LIGO, the axisymmetric distortions are corrected using thermal actuators and CO$_2$ laser projectors, however, non-axisymmetric wavefront distortions remain unmitigated. Deformable mirrors are investigated as a flexible solution for mitigating such defects: by shaping the CO$_2$ beam phase upon reflection, they can imprint the required asymmetric intensity pattern on the lensing optics without introducing frequency dependent noise. The target phase map is computed via a modified Gerchberg-Saxton algorithm. We present simulations of this projection strategy and experimental validation demonstrating consistent reproduction of the desired intensity patterns.

\end{abstract}


\maketitle


\section{Introduction}
The achievable accuracy and sensitivity of precision measurements based on optical interferometry critically depend on the dark fringe contrast. Any residual light at the interferometer output degrades the shot noise limited sensitivity by contaminating the recombined interference pattern. This issue is especially important for high-power gravitational wave (GW) detectors, where imperfect matching of the wavefront and amplitude distributions of the recombined beams due to differential optical aberrations in the two arms produces excess light at the dark port. This is the case for Advanced Virgo \cite{Virgo}, Advanced LIGO \cite{LIGO}, and KAGRA \cite{KAGRA}, whose coordinated operation within the LVK collaboration has enabled the detection of a large number of gravitational wave events, as reported in the gravitational wave transient catalogs (GWTC) \cite{GWTC1, GWTC2, GWTC21, GWTC3, GWTC4, GWTC5}.

Also common mode effects are relevant in GW detectors using optical power recycling. Optical and recycling gains rely on resonant optical cavities whose resonance conditions are preserved through sideband signal extraction techniques \cite{Virgo}. Wavefront distortions---originating from optical surface defects or from thermal lensing caused by laser power absorption mainly driven by mirror coatings \cite{TCS2012, TCS2019})---spoil the resonance condition of the sidebands field in the recycling cavity, thereby reducing detector controllability and sensitivity. Accordingly, deviations from the detector’s ideal optical performance must be constrained within strict requirements to avoid both loss of control and sensitivity degradation. For these reasons, a Thermal Compensation System (TCS) has been developed and commissioned to mitigate optical aberrations and to maintain a stable and optimal operating point of the detector \cite{TCS_AdV, TCS_LIGO}. Looking ahead, the Einstein Telescope high-frequency interferometer will operate with megawatt-level circulating power in the long arms~\cite{ETdesign}, making the thermal aberration control a primary driver of the optical design.

Low-order aberrations are the principal drivers of detector instability. Tilt and uniform optical path offsets can be compensated readily by mirror translation or angular control loops. Consequently, the spherical component---a quadratic term in the optical path length (OPL) distortion---typically represents the dominant uncompensated aberration. The spherical distortion is the leading order of the axisymmetric OPL change produced by laser power absorption in the cavity mirrors.

The standard mitigation approach consists in placing a fused silica compensation plate (CP) in front of the mirror on the recycling cavity side and illuminating it with a shaped CO$_2$ beam that is completely absorbed. The induced thermo-optic and thermo-elastic effects create a divergent aspheric lens complementary to the mirror thermal lensing. The optimal heating profile is computed numerically and implemented experimentally by axisymmetric beam shaping techniques \cite{AdV1_TDR}.

Although the CO$_2$-based correction proved effective in controlling thermal effects in the marginally stable recycling cavity of Advanced Virgo during O3 and O4~\cite{TCS_AdV, TCS_GRASS2019, TCS_GWADW2023}, it is expected to face limitations in future detector upgrades with increased circulating power. The shaped beam profile often exhibits irregular power distributions because of higher-order mode content, imperfections in the shaping optics, and angle-of-incidence effects. Additionally, the interferometer's mirrors exhibit inhomogeneous asymmetric absorption due to imperfections and defects in the coatings, resulting in further wavefront aberrations. These factors introduce non‑axisymmetric OPL distortions that cannot be fully corrected by the symmetric heating pattern and, although negligible in current detectors, will eventually exceed the allowable error budget as the stored power increases. 

The relevance of non-axisymmetric distortions has been recognized since the first generation of detectors. An initial solution was proposed, consisting in a CO$_2$ beam scanning system \cite{Scanning_system} to produce arbitrary heating patterns by rastering a focused spot over a matrix of points. However, such a method potentially introduces coupled noise due to the upconversion of scan frequency and step frequency to the detector sensitivity band through radiation pressure and thermo-optic effect, eventually degrading the sensitivity. An ideal static corrector would solve the issue by depositing a stable, arbitrary asymmetric heating pattern on the CP to produce an optimal asymmetric thermal lens that compensates residual non-axisymmetric wavefront errors without introducing coupled noise.
\begin{figure}[t]
    \centering
    \includegraphics[width=0.8\linewidth]{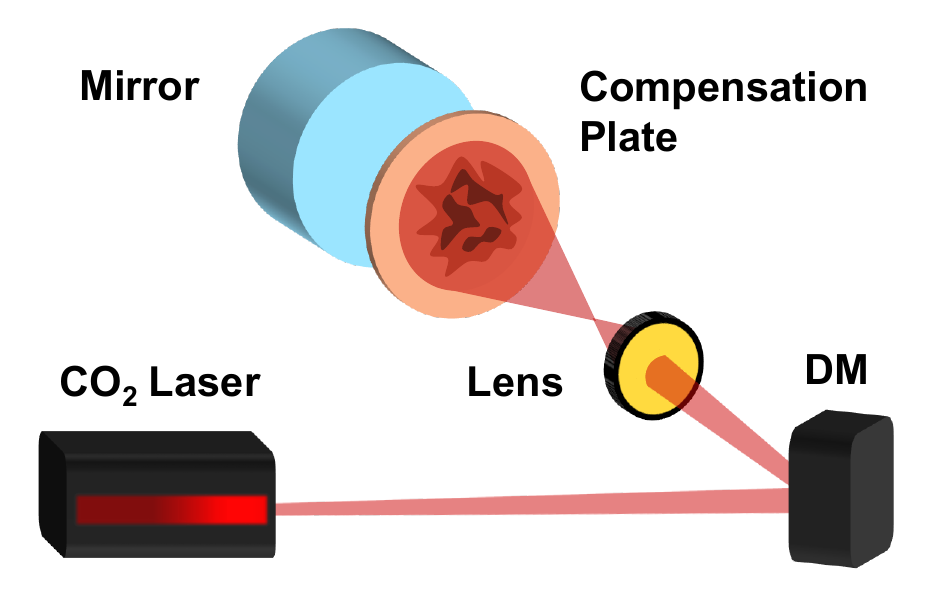}
    \caption{DM's working principle in GW detectors. The DM applies a phase correction to the incident CO$_2$ laser, the resulting non‑axisymmetric intensity profile is projected onto the CP through a lens, introducing a complementary correction to the OPL variations in the mirror \cite{TCS_AdV}.}
    \label{fig:Separated DM actuation}
\end{figure}
A practical static actuator is realized by shaping a CO$_2$ beam in the object plane via phase or intensity modulation, with the resulting image‑plane intensity distribution projected through a lens onto the CP (see Fig.~\ref{fig:Separated DM actuation}). Intensity shaping with static masks lacks flexibility, while liquid-crystal devices are generally unsuitable for high-power beams. For these reasons, phase actuators, and in particular deformable mirrors (DMs), offer the most effective and robust solution. DMs are a mature technology, capable of high-power operation and long-term stability. These instruments are widely used in adaptive optics for astronomy, free space optical communication, medical imaging and optical trapping, among many other applications. Here we consider in particular DM devices based on the continuous faceplate concept. Their relevant features include:
\begin{itemize}
    \item \textit{Actuation architecture}. A thin mirror membrane is deformed by an array of actuators on the rear face. Actuators are designed to be fast and typically operated within a controlled loop to ensure stability.

    \item \textit{Aperture and absorption}. Coating absorption at the CO$_2$ wavelength ($10.6 \mu m$) and the mirror aperture determine the thermal loading; higher absorption requires larger apertures to avoid membrane overheating. Typical apertures are on the order of a few centimeters.

    \item \textit{Spatial-frequency limit}. The number and arrangement of actuators limit the highest spatial frequency that can be reproduced. A rough estimate of the maximum spatial frequency is  $f_{\textit{max}} \sim \frac{1}{2p}$ where $p$ is the actuator pitch, which is the center-to-center distance between adjacent actuators.

    \item \textit{Stroke requirement}. The maximum surface displacement produced by a single actuator (stroke) must be sufficient—typically a fraction of the laser wavelength—to allow an effective beam shaping with a reasonably compact optical layout.
\end{itemize}
\begin{figure}
    \centering
    \includegraphics[width=0.935\linewidth]{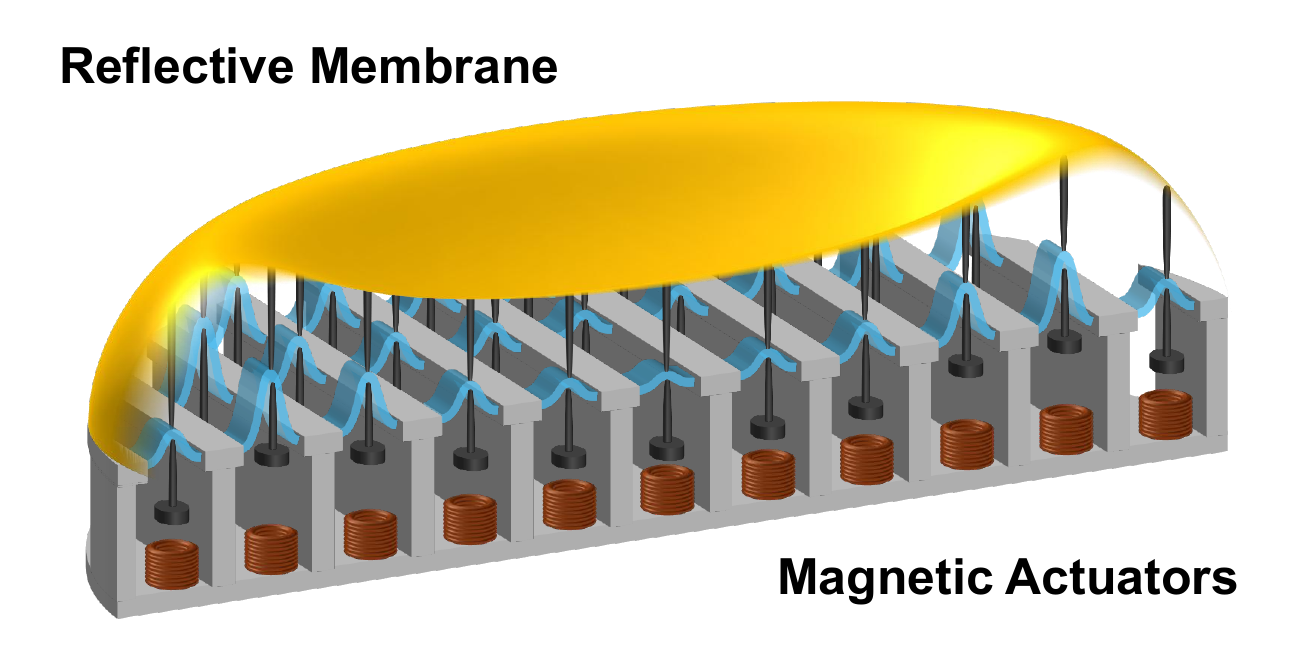}
    \caption{Internal structure of a DM equipped with magnetic actuators: when an actuator is active, electric current runs into its coil, and the magnet above is attracted or repulsed, acting on the reflective surface. Rod springs (blue strips) are used to provide stability and control of the induced displacement.}
    \label{fig:Separated DM structure}
\end{figure}
DMs provide a viable path to implement static, high-power-capable, asymmetric thermal correction with the versatility required for the application in GW interferometers \cite{AdV2_TDR, Aiello_thesis, Taranto_thesis}.

\section{Concepts}

\subsection{Mirror surface actuation}
The standard constructive scheme of a DM includes a set of actuators in contact with the back side of the mirror reflective membrane (see Fig.~\ref{fig:Separated DM structure}). Regardless of the specific actuation mechanism (piezoelectric, electrostatic, or electromagnetic) each actuator is in contact with the rear face of the membrane over a well-defined area and can apply a given amount of force, producing a localized deformation of the surface. The maximum deformation of the mirror membrane is a particular function of the considered actuator and is referred to as \textit{influence function}. Achieving good linearity in the mirror’s response to actuator force generally relies on appropriate design and construction features.

Influence functions are in general approximated by a Gaussian-like expression such as~\cite{Pseudo-Gaussians}:
\begin{equation}
    F_i(r)=\exp \left[\ln (\omega) \left(\frac{r}{p}\right)^{\alpha}\right]
\end{equation}
where $r$ is the distance from the actuator's position, $\omega$ is a parameter representing the amount of coupling toward neighbor actuating points, $p$ is the actuator grid pitch and the Gaussian index $\alpha$ roughly varies within the range $[1.5,2.5]$. However, the actual influence functions depend on the specific mirror technology and design, and in practice their deformation profiles are determined experimentally.

It is usually a good approximation to assume that the deformation $W(x,y)$ resulting from the simultaneous action of the whole set of actuators is given by the direct sum of the $N$ influence functions $F_i(x,y)$ multiplied by their weighting factors $c_i$:
\begin{equation}\label{deformation1}
    W(x,y)=\sum_{i=1}^N c_i F_i(x,y)
\end{equation}
By definition, weighting factors can assume values in the range $[-1,1]$. The actuation of the mirror is governed by control software providing an $N$-dimensional command vector $\vec{c}$. 

The influence functions of a DM are typically characterized through differential measurements of a probe beam’s wavefront, using wavefront sensors such as a Shack–Hartmann (SH) device. The output is a set of two-dimensional discrete maps $F_i (P_{j,k})$ containing $(J\times K)$ values. To write a discretized version of Eq.~\ref{deformation1}, we consider an ordered sequence of the sampling points $P_l$ with a predefined traversal rule $(j,k)\rightarrow (l)$. We define the \textit{influence matrix} (IM), denoted by $M$, as the horizontal concatenation of influence functions:
\begin{equation}
    M_{ij} = F_j (P_i)
\end{equation}
The mirror deformation map obtained by issuing a command vector $\vec{c}$, written in the form of an ordered vector $\vec{W}$, is therefore:
\begin{equation}\label{directsystem}
    \vec{W} = M \cdot \vec{c}
\end{equation}
Under the approximation of linear superposition of actuator effects, the IM fully describes the device operation.

\subsection{Command vector optimization}
In practice, operating a DM requires the device to produce a specific phase imprint, that is, a desired surface deformation. This is typically provided in the form of a discretized two-dimensional map. To get the desired mirror map, we need to determine the command vector values. The computation of $\vec{c}$ would require to invert Eq.~\ref{directsystem}; however, this is normally not feasible since the length $L = J\times K$ of $\vec{W}$ is typically much larger than the actuator number $N$, therefore the IM is far from being square. The linear system in Eq.~\ref{directsystem} is strongly over-constrained. We then search for a command vector that produces a deformed surface as close as possible (finding a least-square solution) to the target $\vec{W}$, so to minimize the norm $\lVert M \cdot \vec{c} - \vec{W} \rVert_2$. This condition is fulfilled by taking the Moore-Penrose pseudo-inverse $(L\times N)$ matrix $M^{-1}$~\cite{Moore, Penrose} and computing:
\begin{equation}
    \vec{c}_{\textit{\scriptsize BEST}} = M^{-1} \cdot \vec{W}
\end{equation}

\subsection{Phase-to-intensity beam shaping: the Modified-Gerchberg-Saxton Approach}
The Rayleigh-Sommerfeld space propagation of optical fields results in a mixing of field intensity and phase. This means that by controlling the phase of a light wave on a given object plane we can get a particular intensity on the image plane. However, in practice, determining the phase change needed to obtain a target intensity is not straightforward.

An efficient phase-retrieval method widely used in the Fraunhofer limit is the Gerchberg-Saxton (G-S) algorithm~\cite{G-S}. The G-S iterative scheme relies on a Fourier-transform propagator, typically implemented through a Cooley–Tukey fast Fourier Transform (FFT). Since in our case the spatial compactness of the DM-based system must be preserved, a pure Fraunhofer-domain propagation cannot be pursued. We therefore replace the Fourier-transform propagator with a Fresnel transfer-function approach, leading to a modified version of the original G-S algorithm (Modified Gerchberg-Saxton or MoG-S~\cite{MoG-S}). 

\subsection{Influence matrix}
\begin{figure*}[t]
    \centering
    \includegraphics[width=0.85\linewidth]{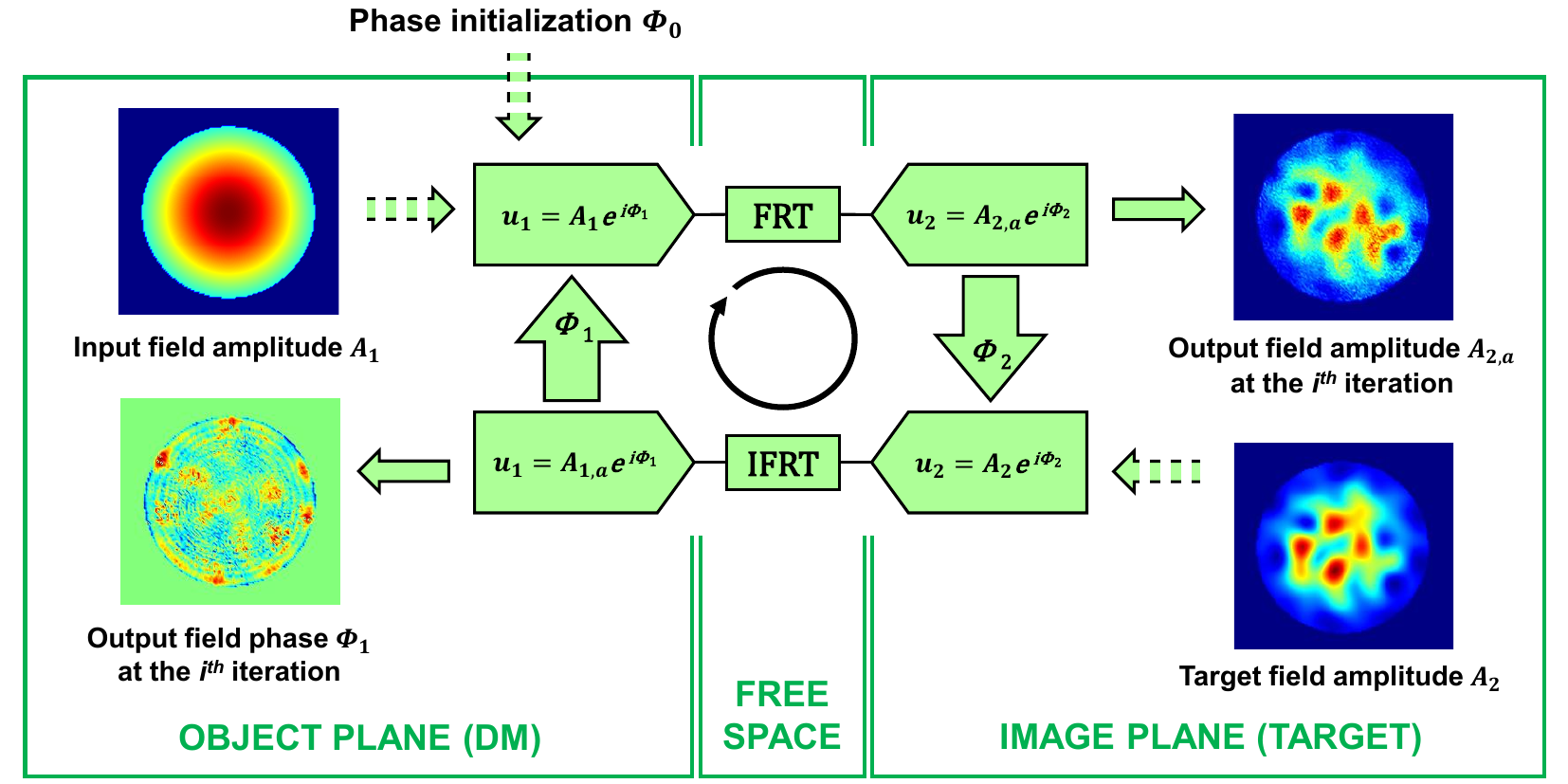}
    \caption{Scheme of the implemented MoG-S algorithm. As in the original G-S scheme, the imaged phase is applied to the target amplitude and propagated back to the object, but, in this case, the free space propagator is a FRT and its inverse. In the scheme, solid arrows refer to fields that are updated at each iteration, while dashed arrows are used to identify input data.}
    \label{fig:MoG-S}
\end{figure*}
Our MoG-S scheme is depicted in Fig.~\ref{fig:MoG-S}. The algorithm requires the initial phase and amplitude of the optical field on the object plane (corresponding to the DM position) to be provided. While the source determines the amplitude, the phase initialization can be chosen to improve the stability and convergence of the iteration\footnote{The actual beam phase on the DM must be subtracted to the output $\Phi_1$ to get the phase correction to apply with the DM. However, in most cases, the beam size on the mirror is large enough that it can be treated as a plane wave.}. Free space propagation to the image plane and back to the object is performed via a Fresnel transform (FRT) and its inverse (IFRT), both implemented using the transfer function approach with an FFT. The original G-S algorithm scheme is otherwise preserved: the imaged phase $\Phi_2$ is applied to the target amplitude $A_2$ and propagated back to the object, where it replaces the input phase. The MoG‑S algorithm operates once the distance between the object and image planes is fixed. In our application, the optimal distance is selected by minimizing the residual difference between the target intensity map and the intensity obtained from the phase map retrieved by MoG‑S (see further, Fig.~\ref{fig:Optimal distance}). With this distance assumed, convergence to a RMS residual error below $6\%$ is usually reached using this MoG-S in $400$ iterations. 

The transfer function of the FRT is defined as follows. 
\begin{equation}\label{eq:FRT TF}
    H(f_X, f_Y) = \exp{\big\{-i \pi \lambda z (f_X^2 + f_Y^2)\big\}}
\end{equation}
where $f_X$ and $f_Y$ are the spatial frequencies along the transverse directions on a plane perpendicular to the propagation vector, while $z$ represents the propagation distance between the planes. Eq.~\ref{eq:FRT TF} implies that, for different wavelengths, the MoG-S algorithm produces identical results, but at an optimal distance that scales according to the following relation between the wavelength and the object–image separation:
\begin{equation}
    \lambda z \approx \textit{constant}
\end{equation}
At the CO$_2$ laser wavelength, the optimal distance is less than $10\textit{cm}$, thereby assuring system compactness.

\section{Deformable mirror characterization}
The aim of this work is to experimentally demonstrate that a DM can reproduce a non-axisymmetric target intensity pattern by applying an appropriate phase correction derived from the developed simulation tools. Two commercial DM devices manufactured by Bertin-Alpao~\cite{ALPAO} were tested: one comprising 97 actuators (DM97) and another one with 192 actuators (DM192). Both devices employ magnetic actuation, in which a thin reflective membrane is suspended above an array of electromagnets, whose local magnetic fields induce controlled deformations of surface. The DM97 provides coarser spatial control of the surface suitable for compensation heating patterns for lower order aberrations, while the DM192 enables higher spatial resolution and more refined shaping.

Magnetic actuation was selected for its key advantages: absence of mechanical contact, which minimizes wear and hysteresis; good linearity across the actuation range; and long-term reproducibility. The reflective membranes are fabricated from materials chosen for low thermal drift and high mechanical stability. Both DMs are suited for high-power CO$_2$ laser beam shaping. They feature a gold-coated reflective membrane that ensures high reflectivity at the CO$_2$ laser wavelength, resistance to oxidation, and the ability to maintain optical performance even at elevated temperatures. The device exhibits a specified damage threshold of $30 W/cm^2$ at the CO$_2$ laser wavelength, ensuring compatibility with thermal compensation applications in gravitational wave detectors. The actuation commands are managed via a dedicated driver, with each actuator controlled by a normalized input value between –1 and 1 (–1 = maximum lowering, 0 = flat configuration, +1 = maximum raising). 
The low hysteresis level ($<2\%$) further contributes to its reproducibility and makes it particularly attractive for reproducible open-loop use. Both models include the high-stability option, which incorporates non-polymer actuator springs and active thermal regulation of the mount, ensuring reproducible deformation and stable performance under open-loop operation. The main properties of DM192 are summarized in Table~\ref{tab:DM192 parameters}.

\renewcommand{\arraystretch}{1.3}
\begin{table}
\centering
\begin{tabular}{c@{\hskip 0.6cm}c}
\hline\hline
\begin{tabular}{@{}c@{}}\text{Parameter}\end{tabular} & 
\begin{tabular}{@{}c@{}}\text{Value}\end{tabular}\\
\hline
\text{Number and class of actuators} & \text{192, magnetic}\\
\text{Diameter} & $21mm$\\
\text{Pitch} & $1.5mm$\\
\text{Coating material} & \text{Gold}\\
\text{Average Stroke} & $4 \mu m$\\
\text{Damage threshold} & $30\textit{W/cm}^2$\\
\hline\hline
\end{tabular}
\caption{Main physical parameters of the DM192 device.}
\label{tab:DM192 parameters}
\end{table}

Operational modes include both static actuation---used to apply steady heating patterns for long-term compensation---and dynamic modulation---employed to follow slow thermal drifts in the optics. Together, these capabilities make the DMs suitable candidates for integration into adaptive thermal compensation systems for gravitational wave detectors.

Both devices were subjected to several tests to assess the operational performances. Linearity and stability are particularly relevant for our scheme of open-loop phase control. The results are summarized below:
\begin{itemize}
    \item[$-$] \textit{Memory effect}. After driving an actuator to +1 and returning it to rest, the residual surface deformation becomes negligible after 2 minutes.
    
    \item[$-$] \textit{Temporal stability}. For a randomly generated phase pattern, the phase correction is stable over time. The RMS difference between phase maps acquired 3.5 hours apart is comparable to the SH sensor noise.
    
    \item[$-$] \textit{Phase actuation}. Selected phase patterns were experimentally reproduced and measured. The RMS difference between the measured and simulated phase maps, converted using the factor $\frac{\lambda}{2\pi}$, is typically $\sim 0.05\mu m$ for an average actuation of $1 \mu m$.
    
    \item[$-$] \textit{Linearity}. The response of each actuator is linear over the limited range [–0.8, +0.8] due to hardware constraints imposed to protect the reflective membrane. Deviations from the expected linear amplitude are below $2\%$.
\end{itemize}

\section{Simulations and Experimental validation}\label{sec:Simulations and Experimental validation}
\begin{figure}[h!]
    \centering
    \includegraphics[width=0.97\linewidth]{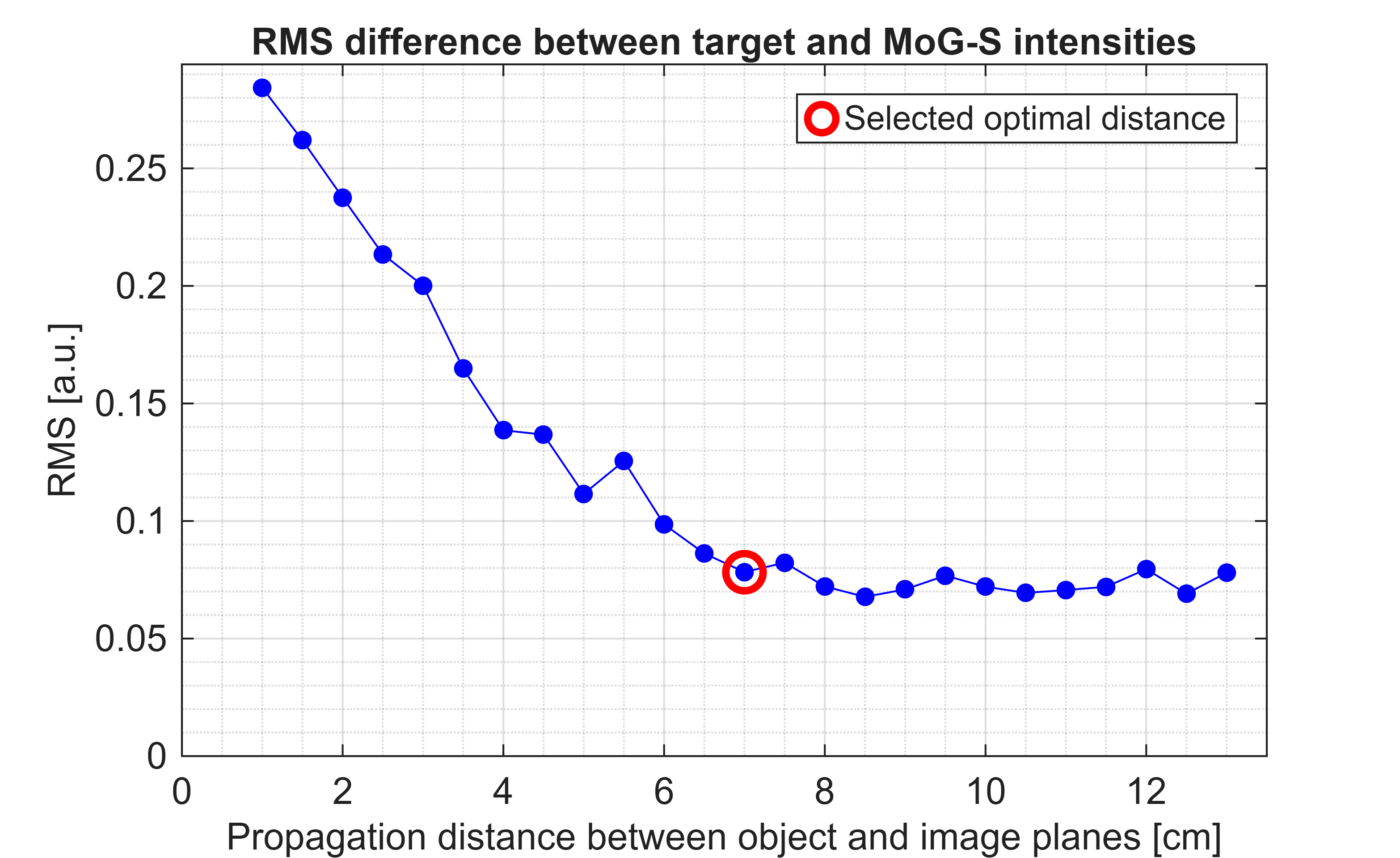}
    \caption{Example of RMS difference between target and MoG-S intensities as a function of the propagation distance from the object plane, using a CO$_2$ laser source. The optimal propagation distance is selected at the onset of the RMS convergence, which corresponds to a trade‑off between intensity reproduction accuracy and layout compactness.}
    \label{fig:Optimal distance}
\end{figure}

The MoG-S algorithm operates on three fundamental inputs: the incident intensity profile at the DM, the target intensity profile to be reconstructed, and the propagation distance between the corresponding planes. In the following, we describe the procedure adopted to define each of these elements prior to applying the algorithm.
\begin{itemize}
    \item[$-$] The \textit{incident intensity} profile on the DM is directly measured at the DM position by a dedicated sensor (see further, Sec.~\ref{sec:Green laser experiment}). This approach yields markedly improved results in experimental validation, compared to using an analytic Gaussian amplitude profile with the beam parameters of the CO$_2$ source.
    
    \item[$-$] A set of randomly generated test \textit{target intensity} maps is prepared to assess the reliability of the method. Different spatial spectra are obtained by filtering out all spatial-frequency components above a specified threshold. To adapt the target to the incident intensity on the DM, which exhibits an almost-Gaussian profile, a sigmoid function is applied to the target. This process, known as apodization, prevents the MoG-S algorithm from reproducing unnecessary high spatial-frequency content at the map edges. It does not represent a limitation, since the peripheral regions---where the incident beam is essentially absent---are irrelevant for the correction.

    \item[$-$] The \textit{propagation distance} between the DM and target planes is a free parameter. Too short distances are known to spoil the MoG-S reconstruction accuracy: this is mostly due to densely distributed discontinuities in the retrieved phase map arising from the restriction of large values to the interval $[-\pi,\pi]$. Conversely, a large $z$ leads to a less compact layout. The optimal value is identified by evaluating the RMS difference between the selected target intensity and the reconstructed intensity obtained after the MoG-S phase correction at varying propagation distances (see Fig.~\ref{fig:Optimal distance}).
\end{itemize}

\begin{figure*}[t]
    \centering
    \includegraphics[width=1\linewidth]{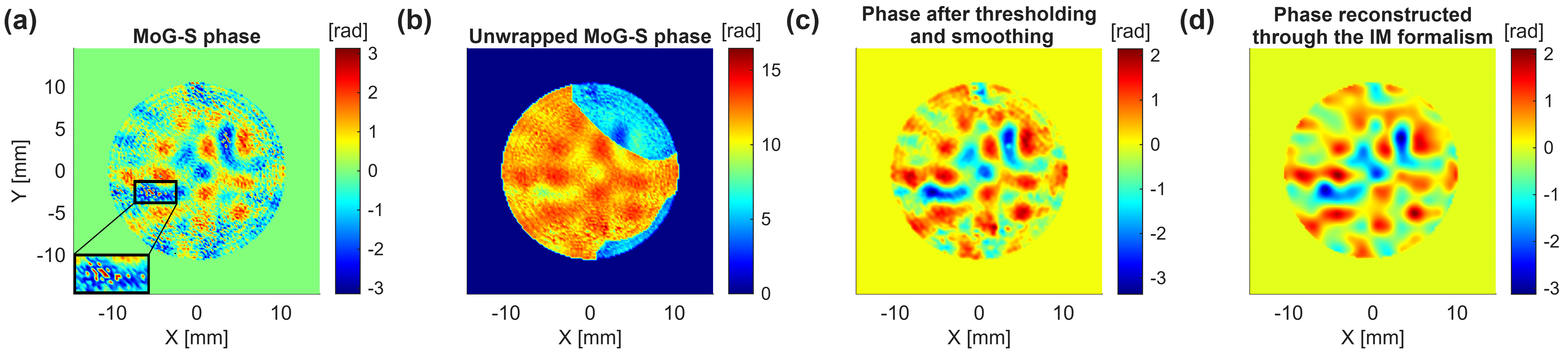}
    \caption{Phase maps illustrating the phase processing. (a) Phase retrieved using the MoG-S algorithm; the inset emphasizes the typical granularity of phase jumps. (b) Unwrapped MoG-S phase obtained via the method proposed by Zhao et al.; non-local discontinuities are still present. (c) Phase after applying a threshold selection to correct the remaining discontinuities, followed by smoothing. (d) Final phase profile obtained for the DM by applying the IM pseudo-inverse method.}
    \label{fig:Phase progression}
\end{figure*}

Once these elements are specified, the MoG-S algorithm computes a candidate phase map to be implemented through the DM. However, the resulting phase distribution typically contains residual discontinuities (phase jumps). Since the DM cannot reproduce discontinuous regions, these must be removed through a phase unwrapping procedure. Among the various unwrapping strategies evaluated, the method proposed by Zhao et al.~\cite{Unwrap} provided the most reliable performance for phase patterns derived from the MoG-S algorithm. The application of the selected unwrapping algorithm typically reduces the distributed map jumps to few discontinuity lines. These residual artifacts can be effectively removed by applying a simple threshold-driven correction. The resulting phase map is subsequently smoothed using a two-dimensional Gaussian kernel with a standard deviation of 1, thereby suppressing high-frequency artifacts and ensuring a more physically consistent representation.

We consistently observed that the density of phase jumps increases as the spatial frequency of the target intensity decreases. This is a classic case of ill-posedness, induced by the non-uniqueness of the solution retrieved by the MoG-S algorithm. Phase discontinuities naturally arise in G-S–type algorithms due to their local projection scheme and the lack of smoothness regularization terms. This situation corresponds to a large degeneracy in the solution space. However, when the target intensity contains significant high spatial-frequency components, these discontinuities are limited by intensity constraints, thereby strongly suppressing their occurrence. Phase jumps are therefore expected to severely impact phase retrieval when the intensity map is smooth. As a result, the unwrapping algorithm does not work optimally. The non-uniqueness of the solution represents the main drawback of the MoG-S algorithm: the number of all possible solutions increases with the number of pixels in the maps, and the particular result depends on the input phase $\Phi_0$.

At this stage, the candidate phase map to be reproduced by the DM is available. The IM formalism is then employed to determine the actuator weights and to simulate the phase map that the DM can realistically reproduce. Fig.~\ref{fig:Phase progression} summarizes the sequence of processes leading to the final DM phase map. Finally, the resulting intensity profile at the image plane is simulated by propagating the optical field after the DM phase correction.

The subsequent step consists in the experimental validation of the simulation results. An initial test was performed with a green laser, allowing direct monitoring of the phase correction applied by the DM through a dedicated SH sensor. The procedure was then repeated using a CO$_2$ laser though in this case without phase monitoring as our setup does not include wavefront sensors for this wavelength.

\subsection{Green laser experiment}\label{sec:Green laser experiment}
\begin{figure}[h!]
    \centering
    \includegraphics[width=0.90\linewidth]{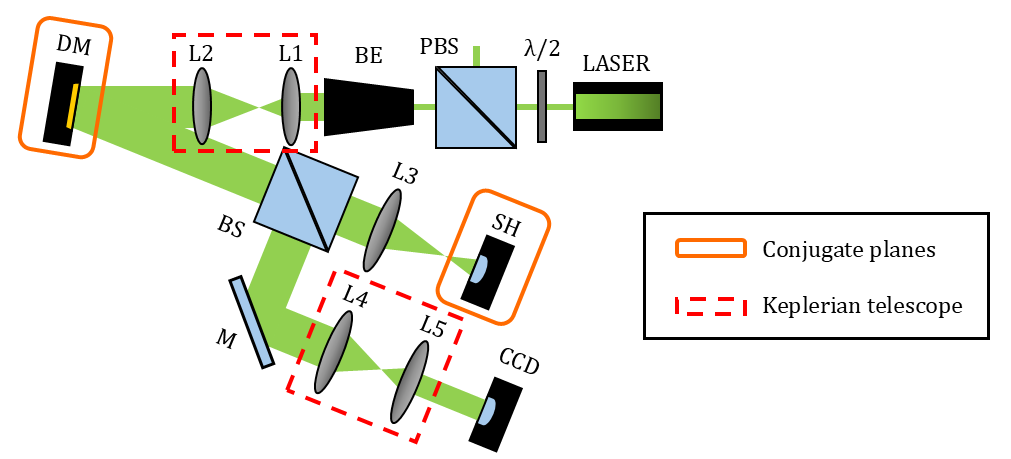}
    \caption{Optical layout based on a green laser, assembled to measure the intensity map resulting from the phase correction applied by the DM, which is simultaneously monitored using a SH sensor positioned in the DM conjugate plane through the $L_3$ lens. The optical elements are labeled as follows: PBS (polarized beam splitter), BE (beam expander), L (lens), BS (beam splitter), M (mirror). The combined use of a half-wave plate and a PBS allows the control of image contrast on the CCD.}
    \label{fig:Green optical layout}
\end{figure}

The use of a green laser ($\lambda = 532nm$) enabled a preliminary check of the experimental process in a controlled environment, where laser propagation could be visually inspected and the DM phase imprint was directly determined by the SH sensor. The assembled optical layout is based on the following key elements (see Fig.~\ref{fig:Green optical layout}):
\begin{itemize}
    \item[$-$] The laser beam is significantly expanded through a beam expander BE and the lenses $L_1$ and $L_2$ to ensure uniform intensity and a regular wavefront upon incidence on the DM. These conditions are essential for achieving improved corrections, as predicted by simulations. The beam width should not exceed the DM diameter.

    \item[$-$] The DM is imaged onto the SH plane through the projection lens $L_3$ to measure the applied phase correction.

    \item[$-$] The target intensity map plane is imaged onto a CCD by using  two lenses L4 and L5, in order to fit the beam width to the CCD acceptance. In addition, the CCD can be employed to measure the incident intensity on the DM, to be used as input for the simulations.
\end{itemize}
At the green laser wavelength, the optimal propagation distance between DM and image plane is on the order of

\begin{figure*}[t]
    \centering
    \includegraphics[width=0.85\linewidth]{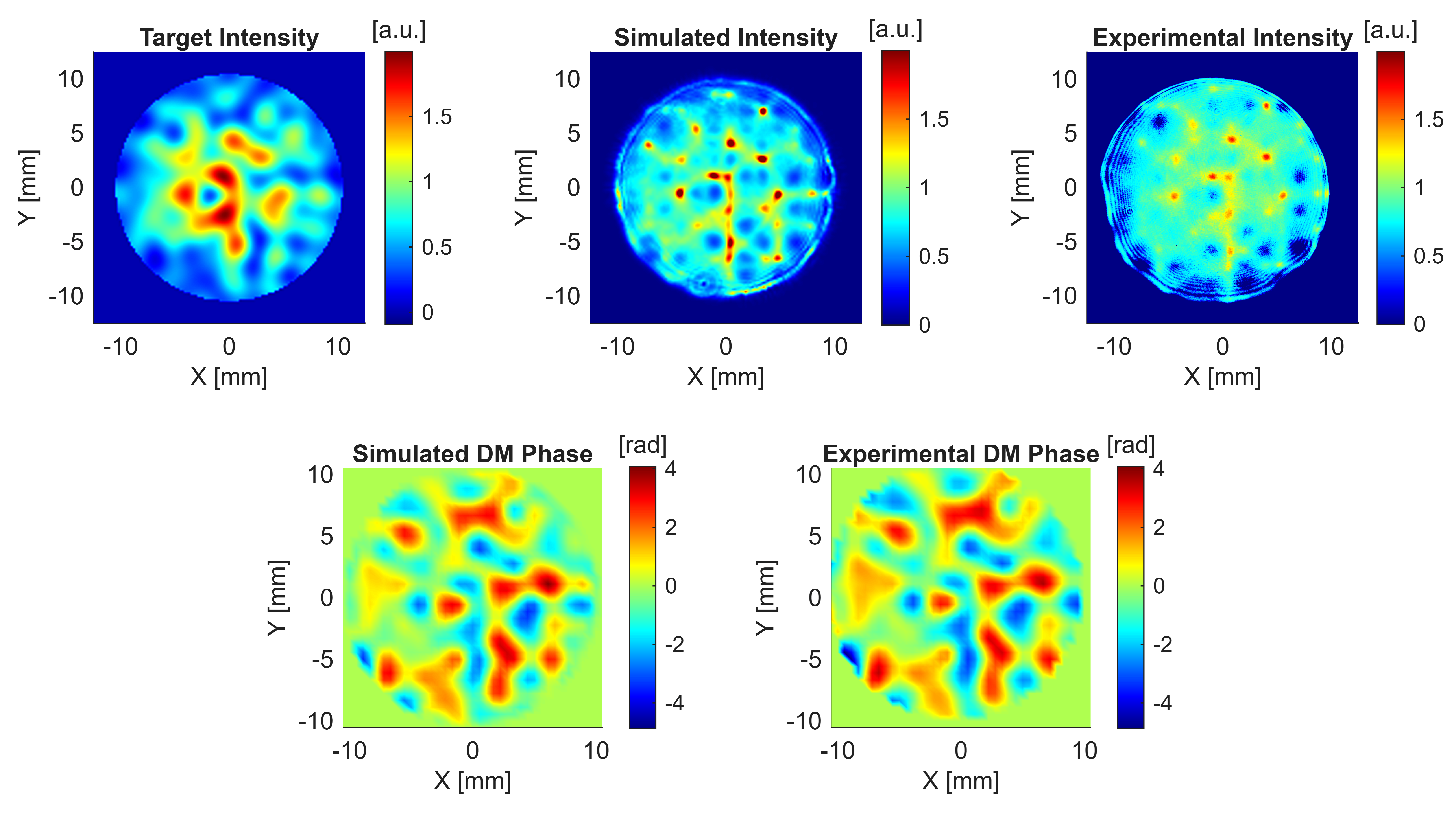}
    \caption{Top row: Target, simulated, and experimental intensity maps for a target with spatial frequencies up to $f_{\textit{\footnotesize pitch}} / 2$ in the green laser experiment. The RMS differences, normalized to the RMS of the target intensity, are approximately $33\%$ between the experimental and simulated maps, and $35\%$ between the experimental and target maps. Bottom row: Simulated and experimental phase maps. The DM provides the expected phase correction, with an RMS difference of about $7\%$ of the full phase range.}
    \label{fig:Intensity and phase comparison green laser}
\end{figure*}
\begin{figure*}[t]
    \centering
    \includegraphics[width=0.85\linewidth]{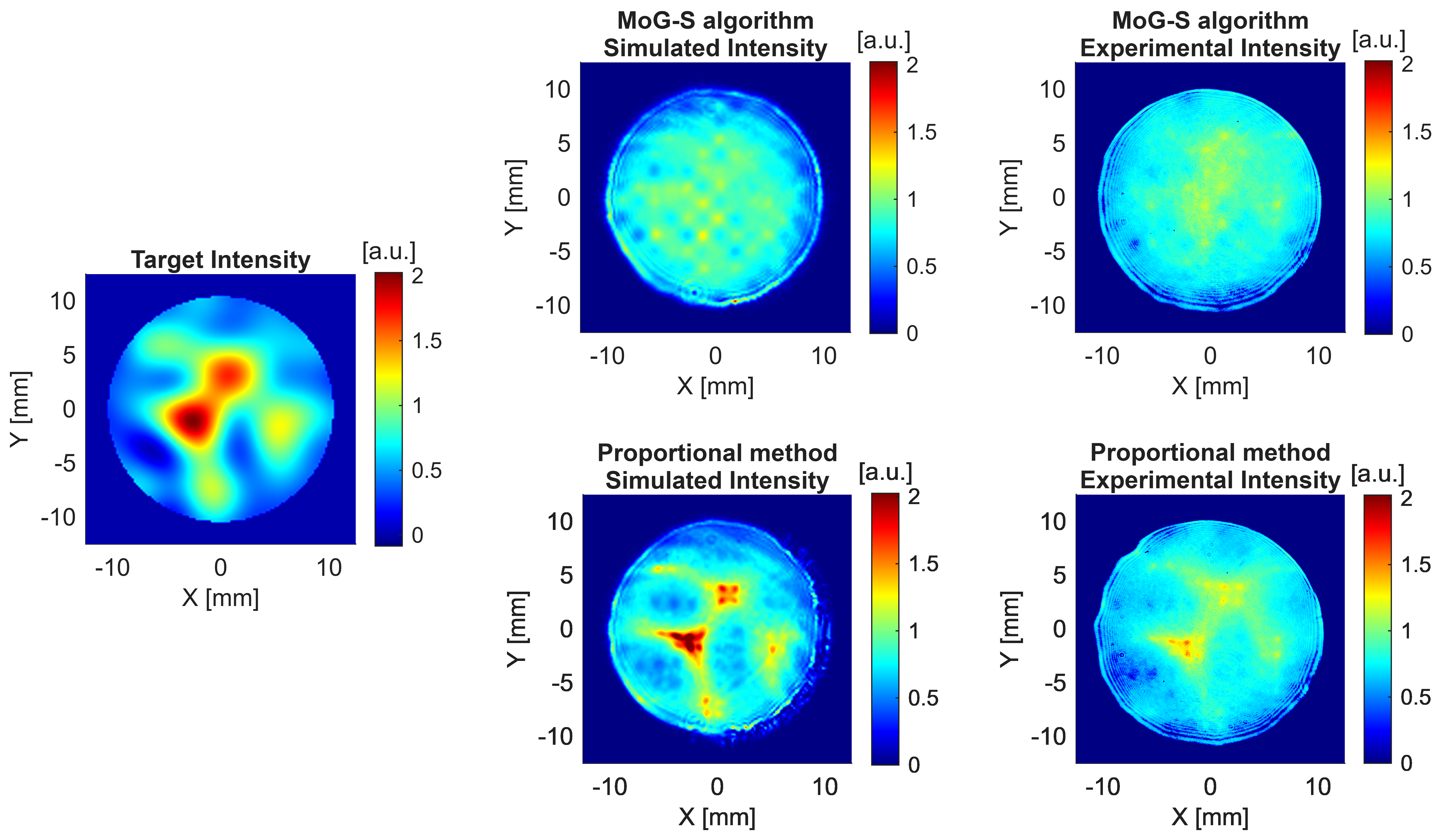}
    \caption{Comparison of simulated and experimental intensity maps for two phase reconstruction methods. Left: Target intensity map randomly generated with spatial frequencies up to $f_{\textit{\footnotesize pitch}} / 4$. Top row: Simulated and experimental intensity maps obtained using the MoG-S algorithm. The RMS difference, normalized to the RMS of the target, between the target and experimental maps is approximately $37\%$. Bottom row: Simulated and experimental intensity maps obtained using the proportional method. The RMS difference, normalized to the RMS of the target, between the target and experimental maps is reduced to $31\%$, indicating improved fidelity compared to the MoG‑S approach.}
    \label{fig:Proportional phase method green laser}
\end{figure*}
\clearpage
\onecolumngrid
\renewcommand{\arraystretch}{1.3}
\begin{table*}[ht]
\centering
\begin{tabular}{c@{\hskip 0.6cm}c@{\hskip 0.6cm}c@{\hskip 0.6cm}c}
\hline\hline
\begin{tabular}{@{}c@{}}\text{Method to derive}\\[-5pt]\text{phase correction}\end{tabular} & 
\begin{tabular}{@{}c@{}}\text{Maximum spatial frequency}\\[-5pt]\text{in the target map}\end{tabular} & 
\begin{tabular}{@{}c@{}}\text{RMS difference of Simulated and}\\[-5pt]\text{Experimental intensities [\%]}\end{tabular} & 
\begin{tabular}{@{}c@{}}\text{RMS difference of Target and}\\[-5pt]\text{Experimental intensities [\%]}\end{tabular} \\
\hline
MoG-S & $f_{\textit{\small pitch}}/2$   & 33 & 35 \\
MoG-S & $f_{\textit{\small pitch}}/2.5$ & 27  & 34 \\
MoG-S & $f_{\textit{\small pitch}}/3$   & 24 & 34 \\
MoG-S & $f_{\textit{\small pitch}}/4$   & 23 & 37 \\
Proportional phase & $f_{\textit{\small pitch}}/4$ & 29 & 31 \\
\hline\hline
\end{tabular}
\caption{Comparison of the RMS difference between the simulated and experimental intensity maps, and between the target and experimental profiles, both expressed as a percentage relative to the RMS of the corresponding target intensity map for each case considered in the green laser analysis.}
\label{tab:Green laser results}
\end{table*}
\twocolumngrid
\hspace{-0.35cm}$1m$. Following the procedure previously described, several target intensity maps were randomly generated up to a spatial frequency limit expressed as fractions of the pitch frequency of the DM ($f_{\textit{\small pitch}} \equiv \frac{1}{p} \approx 666m^{-1}$), from $f_{\textit{\small pitch}} / 2$ to $f_{\textit{\small pitch}} / 4$. For each case, the intensity map obtained after DM actuation was simulated through the IM formalism and the corresponding experimental map was measured. The RMS difference between the two maps was adopted as the figure of merit to assess the reliability of the method. The results obtained for the selected intensity patterns at different spatial frequencies are reported in Table~\ref{tab:Green laser results}. Fig.~\ref{fig:Intensity and phase comparison green laser} illustrates a representative case at spatial frequency $f_{\textit{\small pitch}} / 2$. In all the tested cases, the experimental intensity exhibits good agreement with the pattern predicted by the simulated actuation, while retaining the principal features and overall trends of the target map. The measured phase actuation is consistent with the simulated map, obtained through the IM formalism, generally with an RMS difference less than $10\%$ of the
full range. All maps are normalized with respect to the
integrated total power, and a cross-correlation procedure was implemented to properly align the experimental and the simulated maps.

As the spatial frequency of the selected target decreases, discrepancies become more pronounced. This behavior is expected, as discussed in Sec.~\ref{sec:Simulations and Experimental validation}: lower target frequencies yield increasingly discontinuous phase reconstructions by the MoG‑S algorithm, leading to larger deviations between the target and simulated intensities (see Fig.~\ref{fig:Proportional phase method green laser}).

To address the problem of solution degeneracy at low spatial frequencies, we propose an alternative approach based on implementing a phase map directly proportional to the target intensity rather than relying on the MoG‑S retrieval. Empirically, the phase recovered by MoG‑S often resembles the target intensity, suggesting that this pragmatic, though not theoretically rigorous, strategy can provide a viable substitute when the MoG‑S solution is limited. The phase map is defined to be proportional to
the target intensity:
\begin{equation}\label{eq:Proportional phase}
    \Delta \phi = \alpha \hspace{0.05cm} \Bigg( \frac{\text{max}[I_{\textit{\small target}}] - \text{min}[I_{\textit{\small target}}]}{2} - I_{\textit{\small target}} \Bigg)
\end{equation}
where max[...] and min[...] indicate the maximum and minimum value, respectively, and the gain $\alpha$ is a free parameter. This function describes a phase variation proportional to a rescaled version of the target intensity $I_{\textit{\small target}}$. The factor $\alpha$ is optimized by minimizing the RMS difference between the target intensity and the intensity resulting from the application of the phase map described by Eq.~\ref{eq:Proportional phase}. Fig.~\ref{fig:Proportional phase method green laser} shows the results obtained from this analysis applied to the lowest frequency case, with spatial frequency $f_{\textit{\small pitch}} / 4$, compared to the MoG-S solution. The resulting RMS difference between experimental and target profiles obtained with the phase map proportional to the target ($31\%$ of the RMS of the target map), is smaller than that achieved using the MoG‑S algorithm ($37\%$ of the RMS of the target map), indicating improved performance over the MoG‑S approach.

\subsection{CO$_2$ laser experiment}
We now present the same analysis performed using a CO$_2$ laser source. Compared to the green laser setup,

\begin{figure*}[t]
    \centering
    \includegraphics[width=0.85\linewidth]{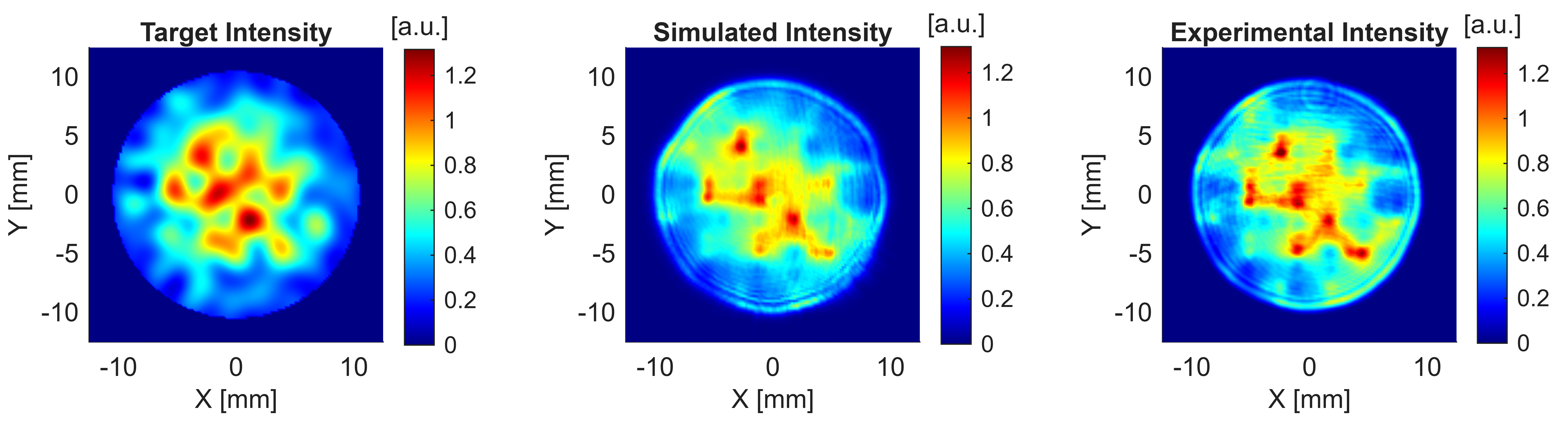}
    \caption{Target, simulated, and experimental intensity maps for a target with spatial frequencies up to $f_{\textit{\footnotesize pitch}} / 2$ in the CO$_2$ laser experiment. The RMS differences, normalized to the RMS of the target intensity, are approximately $18\%$ between the experimental and simulated maps, and $26\%$ between the experimental and target maps. This indicates that there remains room for improvement in reproducing the target distribution.}
    \label{fig:Intensity comparison CO2 laser}
\end{figure*}
\begin{figure*}[t]
    \centering
    \includegraphics[width=0.85\linewidth]{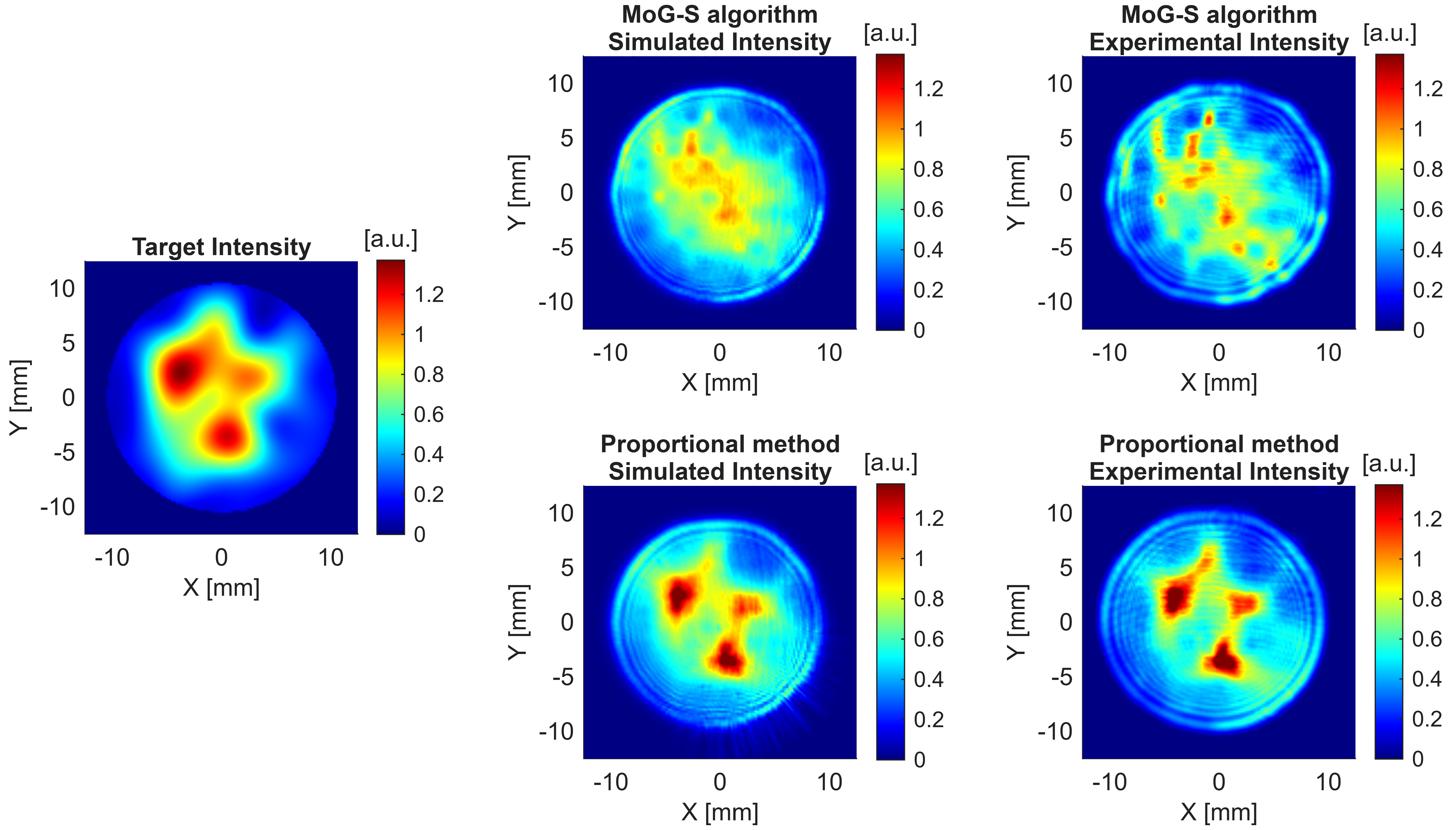}
    \caption{Comparison of simulated and experimental intensity maps for two phase reconstruction methods in the CO$_2$ laser experiment. Left: Target intensity map randomly generated with spatial frequencies up to $f_{\textit{\footnotesize pitch}} / 4$. Top row: Simulated and experimental intensity maps obtained using the MoG-S algorithm. The RMS difference, normalized to the RMS of the target, between the target and experimental maps is approximately $39\%$. Bottom row: Simulated and experimental intensity maps obtained using the proportional method. The RMS difference, normalized to the RMS of the target, between the target and experimental maps is reduced to $31\%$, indicating improved fidelity compared to the MoG‑S approach.}
    \label{fig:Proportional phase method CO2 laser}
\end{figure*}

\begin{figure}[h!]
    \centering
    \includegraphics[width=0.85\linewidth]{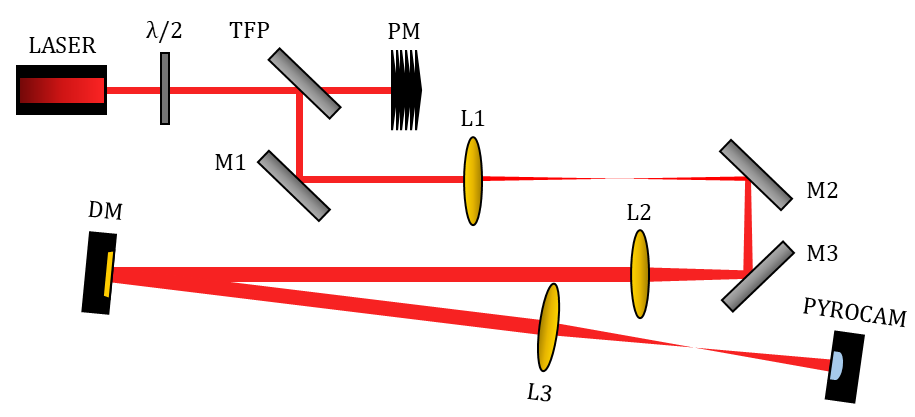}
    \caption{Optical layout based on a CO$_2$ laser, assembled to measure the intensity map resulting from the phase correction applied by the DM. The optical elements are labeled as follows: TFP (thin film polarizer), PM (powermeter), M (mirror), L (lens).}
    \label{fig:CO2 optical layout}
\end{figure}

\renewcommand{\arraystretch}{1.3}
\begin{table*}[t]
\centering
\begin{tabular}{c@{\hskip 0.6cm}c@{\hskip 0.6cm}c@{\hskip 0.6cm}c}
\hline\hline
\begin{tabular}{@{}c@{}}\text{Method to derive}\\[-5pt]\text{phase correction}\end{tabular} & 
\begin{tabular}{@{}c@{}}\text{Maximum spatial frequency}\\[-5pt]\text{in the target map}\end{tabular} & 
\begin{tabular}{@{}c@{}}\text{RMS difference of Simulated and}\\[-5pt]\text{Experimental intensities [\%]}\end{tabular} & 
\begin{tabular}{@{}c@{}}\text{RMS difference of Target and}\\[-5pt]\text{Experimental intensities [\%]}\end{tabular} \\
\hline
MoG-S & $f_{\textit{\small pitch}}/2$   & 18 & 26 \\
MoG-S & $f_{\textit{\small pitch}}/2.5$ & 17  & 31 \\
MoG-S & $f_{\textit{\small pitch}}/3$   & 15 & 34 \\
MoG-S & $f_{\textit{\small pitch}}/4$   & 20 & 39 \\
Proportional phase & $f_{\textit{\small pitch}}/4$ & 20 & 31 \\
\hline\hline
\end{tabular}
\caption{Comparison of the RMS difference between the simulated and experimental intensity maps, and between the target and experimental profiles, both expressed as a percentage relative to the RMS of the corresponding target intensity map for each case considered in the CO$_2$ laser analysis.}
\label{tab:CO2 laser results}
\end{table*}
\clearpage
\hspace{-0.35cm}the CO$_2$-laser-based system requires stronger DM actuation to achieve the same phase correction, because of the larger wavelength, and the optimal target intensity map is found at a shorter propagation distance (on the order of $5 cm$), due to the scaling relation between wavelength $\lambda$ and propagation distance $z$
described by Equation~\ref{eq:FRT TF}. The optical layout retains the same fundamental components as the green laser system (see Fig. \ref{fig:CO2 optical layout}), with the exceptions that the CCD is replaced by a pyroelectric sensor camera (hereafter referred to as pyrocam) and no wavefront sensors are available. As previously, the analysis was carried out using different intensity patterns at spatial frequencies ranging from $f_{\textit{\small pitch}} / 2$ to $f_{\textit{\small pitch}} / 4$, with results summarized in Table~\ref{tab:CO2 laser results}. Fig.~\ref{fig:Intensity comparison CO2 laser} illustrates the case of $f_{\textit{\small pitch}} / 2$.
In all cases, the experimental data reproduce the main features of the target intensity map, particularly at higher frequencies, showing good agreement with the simulated correction. At lower frequencies, as expected, the correction progressively degrades. Therefore, we applied the method described by Eq.~\ref{eq:Proportional phase}, to assess whether it improves performance (see Fig.~\ref{fig:Proportional phase method CO2 laser}). The resulting RMS difference between experimental and target profiles obtained with the proportional phase map ($31\%$ of the RMS of the target map) is smaller than that achieved with the MoG‑S algorithm ($39\%$ of the RMS of the target map), confirming improved performance relative to MoG‑S.

\section{Conclusions and future work}
This study demonstrates that deformable mirrors can effectively project stable heating patterns for correcting for non-axisymmetric optical aberrations in the mirrors of a gravitational wave interferometer. The target intensity distribution is achieved by applying to the CO$_2$ beam a phase profile obtained by exploiting a modified Gerchberg-Saxton algorithm, and an intensity-proportional method.

A preliminary investigation with a green laser was carried out to validate the simulated results, identify potential limitations, and monitor the phase correction induced by the DM. Subsequently, a CO$_2$ laser experiment confirmed that the DM effectively reproduces the simulated intensity pattern, obtained via the influence matrix formalism, with an average RMS deviation of approximately $16-17\%$ relative to the RMS of the target map. In contrast, when comparing the experimental map directly to the target, the RMS difference increases to about $30-31\%$, nearly twice the previous value. This leaves room for further improvement, both in the phase retrieval pipeline and in the experimental implementation through the deformable mirrors. By eye inspection, the correction applied by the DM successfully captures the global features of the target. We plan to assess the compensation performance of the obtained profiles via a real life case test, by simulating the impact of the pattern projection on the operation of a full Advanced Virgo model, looking at the effects in terms of optical path length correction, higher order modes budget, carrier and sidebands gain.

A newly proposed method, based on the assumption of a phase map proportional to the target intensity, demonstrates better performances over the MoG‑S algorithm at low spatial frequencies. This method is inherently suited to the implementation of a closed-loop projection system, in which the discrepancy between the target and experimentally obtained profiles on the image plane is employed as an error signal for the optimization of the heating pattern.

\begin{acknowledgments}
This work was supported by the Italian National Recovery and Resilience Plan (PNRR), Mission 4 “Education and Research,” Component 2 “From Research to Business,” under the project ETIC – Einstein Telescope Infrastructure Consortium. The authors acknowledge the support of the Italian Ministry of University and Research (MUR) and the European Union – NextGenerationEU. The deformable mirror DM192 used in this study was purchased with funds from the PNRR ETIC project.
\end{acknowledgments}

\bibliography{references.bib}

\end{document}